\documentstyle[aps,prl,epsf,preprint]{revtex}
\def\la{\mathrel{\mathpalette\fun <}}
\def\ga{\mathrel{\mathpalette\fun >}}
\def\fun#1#2{\lower3.6pt\vbox{\baselineskip0pt\lineskip.9pt
        \ialign{$\mathsurround=0pt#1\hfill##\hfil$\crcr#2\crcr\sim\crcr}}}

\begin{document}

\vspace{0.5in}

{\title{\vskip-2truecm{\hfill {{\small  SISSA/AP/95/117\\
        \hfill hep-ph/yymmnn\\
        }}\vskip 1truecm} {\bf Towards a Nonequilibrium  Quantum Field
Theory  }\vskip 0.1truecm {\bf Approach to Electroweak Baryogenesis}}

\vspace{2cm}

\author{ANTONIO RIOTTO\thanks{riotto@tsmi19.sissa.it. Address after
November 95: Theoretical Astrophysics Group, NASA/Fermilab, Batavia,
IL60510, USA.}}

\vspace{1.0cm}

\address{International School for Advanced Studies, SISSA-ISAS,\\
Strada Costiera 11, I-34014, Miramare, Trieste, Italy\\and\\
Istituto Nazionale di Fisica Nucleare, Sezione di Padova, 35100 Padova,
Italy}

\maketitle

\vspace{1.cm}

\begin{abstract}
\baselineskip 16pt

We propose a general method to compute $CP$-violating observables
from extensions of the standard model in the
context of electroweak baryogenesis. It is alternative to the
one recently developed by Huet and Nelson and
relies on a nonequilibrium quantum field theory approach. The method
is valid for all shapes and sizes
of the bubble wall expanding in the thermal bath during a first-order
electroweak phase transition. The quantum physics of $CP$-violation and
its suppression coming from the incoherent nature of thermal processes
are also made explicit.

\end{abstract}

\thispagestyle{empty}

\newpage
\pagestyle{plain}
\setcounter{page}{1}
\def\beq{\begin{equation}}
\def\eeq{\end{equation}}
\def\beqa{\begin{eqnarray}}
\def\eeqa{\end{eqnarray}}
\def\tr{{\rm tr}}
\def\x{{\bf x}}
\def\p{{\bf p}}
\def\k{{\bf k}}
\def\z{{\bf z}}
\baselineskip 20pt
\begin{flushleft}
{\large\bf 1. Introduction}
\end{flushleft}
\vspace{0.5cm}

It is commonly believed  that many of the current questions
of cosmology can be answered by studying the
nontrivial dynamics of the approach to equilibrium in complex systems.
Nevertheless, despite their immense relevance, only very recently
much effort has been made
to understand nonequilibrium phenomena occurred in the early Universe.
For instance, recent
investigations on the non-linear quantum dynamics of scalar fields
have implications for the reheating after the inflationary era
and reveal that particle
production appears to be significantly different from linear estimates
due to the time evolution of the inflation field \cite{noneq}. The
quantum non-linear effects lead to an extremely effective dissipational
dynamics and particle production even in the simplest self-interacting
theory where the single particle decay is kinematically forbidden.

At the electroweak scale, the focus has been in generation of the baryon
asymmetry during a first-order phase transition where
the $SU(2)_L\otimes U(1)_Y$ symmetry  is broken down to
$U(1)_{{\rm em}}$ \cite{ew}. Even though there are certain questions
related to the reliability of the perturbative expansion for weak enough
transitions \cite{quiros}, it is well established that nonequilibrium
conditions are a crucial ingredients for baryogenesis.

An interesting attempt to explain the observed baryon to entropy ratio
$n_B/s\sim 10^{-10}$ was made by Farrar
and Shaposhnikov \cite{fs} in the context of Standard
Model (SM). However, it was later shown  \cite{huet}
that the $CP$-violating processes of quantum interference provided by
the Kobayashi-Maskawa matrix were far too slow in comparison with the
rapid quark-gluon plasma interactions responsible for the loss of
quantistic coherence.
In contrast, many extensions of the standard model, like the Minimal
Supersymmetric Standard Model,
contain additional sources of $CP$ in the Higgs sector (which requires
at least two Higgs doublets) and can account for the
observed baryon asymmetry.

Both in the case of explicit \cite{ex1} or
spontaneous \cite{noi} $CP$-violation, particle mass matrices acquire a
nontrivial space-time dependence when bubbles of the broken phase
nucleate and expand during a first-order electroweak phase
transition. This space-time dependence cannot be rotated away at two
adjacent points by the same unitary transformation and gives rise to
sufficiently
fast nonequilibrium $CP$-violating  effects inside the wall of a
bubble of broken phase expanding in the plasma.

Two different limits were originally investigated in the literature. In the
case of thin bubble walls, characterized by the condition
$\tau/L_{{\rm w}}\gg 1$, where $L_{{\rm w}}$
and $\tau$ are the wall thickness and the
typical particle interaction time, the asymmetric (in fermion numbers)
reflection of
particles off the bubble wall is the dominant effect \cite{thin}.
The induced fermion
number flux is then reprocessed into a baryon asymmetry by the anomalous
$(B+L)$-violating sphaleronic transitions in the unbroken phase
\cite{sp}.

In the opposite limit of thick bubble walls, $\tau/L_{{\rm w}}\ll 1$, local
operators coupling the baryon current, or related currents, to the
space-time derivative of the $CP$-violating phase act as classical
$CP$-violating
perturbations inside the bubble wall \cite{thick,thicknoi} and
effective chemical potentials account for particle asymmetries.

In both scenarios these $CP$-violating sources are
locally induced by the passage of the wall and fuel baryogenesis. They
are then inserted  into a set  of classical Boltzmann equations for particle
distribution densities which permit to describe
Debye screening  of induced
gauge charges \cite{deb}, particle number changing reactions \cite{cha} and
to trace the crucial role
played by diffusion \cite{tra}.
Indeed, it has been convincingly established that
transport allow  $CP$-violating charges to  efficiently
diffuse in
front of the advancing bubble wall where anomalous electroweak baryon
violating processes are unsuppressed, thus greatly enhancing the final
baryon asymmetry.

A new method to compute the effects of $CP$-violation from
extensions of SM on the electroweak baryogenesis has been recently
developed by Huet and Nelson \cite{newmethod1} and subsequently
applied to supersymmetric baryogenesis \cite{newmethod2}. It
takes into account both the effects of scattering from thermal particles
and the terms which lead to $CP$-violation in particle propagation,
reflecting the interplay between the coherent phenomenon of
$CP$-violation and the incoherent nature of plasma physics. It is also
valid for generic wall shapes and sizes with a significant improvement
over the previous estimates. As a matter of fact,
particle interaction times are
neither very small nor very large compared to the thickness of the wall
and the thin (thick) wall limit can overestimate (underestimate) the
amount of baryon asymmetry. Moreover, particles moving with large
oblique angles relative to the advancing wall are likely to interact
many times when inside the wall and the thin wall limit may not be
applied.

When taking the thick wall limit,
the method predicts the correct
behaviour with the mass background in agreement with the results of ref.
\cite{thicknoi} and evades
the assumption that local particle distributions relax towards thermal
equilibrium according to some classical equations of motion, which is
not well-grounded for particles whose wavelength perpendicular to the
wall is larger than their mean free path and makes the classical treatment
of $CP$-violation not adequate.

Technically speaking,  in refs. \cite{newmethod1,newmethod2}
$CP$-violating
sources were described in terms of quantum mechanical $CP$-violating
reflection and transmission from layers of the phase boundary.
Thermalization effects  on particle distributions are included by
averaging over a layer of
thickness equal to the free mean path $\Delta=\tau v$ ($v$ being the particle
velocity) the  currents
$J_{\pm}$ in the rest frame of the wall and
resulting from particles moving toward the positive (negative)
direction $z$ perpendicular to the wall. For example, the current $J_{+}$
receives contributions from either particles
originating from the thermal bath at a certain point $z$ an moving
with a positive velocity and transmitted to $z+\Delta$ (with the
possibility of different scatterings along the path), and from
particles originating at $z+\Delta$ and moving with velocity
$-v$ and being reflected back towards $z+\Delta$.  When boosted to the
plasma frame, these currents give rise
 to the $CP$-violating
source terms  for the final baryon asymmetry.

It is indisputable that the ultimate answer on electroweak baryogenesis can
be provided only by a self-consistent nonequilibrium Quantum Field Theory
(QFT)
approach. Kinetic theory and classical Boltzmann equations have been used to
describe the dynamics of particles treated as classical with a defined
position, energy and momentum. However, the validity of the kinetic
theory is restricted by the condition that the mean free path of
particles must be larger than any other microscopic length scale. In
particular, the mean free path must be large compared to the Compton
wavelength of the underlying particle in order for the classical picture
to be valid, which is not guaranteed for
particles with a small momentum perpendicular to the
wall. Moreover, since the effects of scattering on particle propagation can be
accounted for by substituting particles with quasiparticles with a
modified dispersion relation and a
given damping rate $\Gamma$ proportional to the imaginary part
of the self-energy, only when the energy of quasiparticles is large
compared to $\Gamma$ it is possible to speak about coherent excited
states and describe them through classical Boltzmann equations
opportunely modified to include plasma effects. \cite{jeon}.
In the opposite limit, the key elements
to address the space-time evolution of distribution functions are
provided by the Wigner function techniques  which date back
to Wigner's work on transport phenomena \cite{wigner}.
Quantum distribution functions are the correct
functions to describe particles in an interacting, many-particle
environment and obey the quantum Boltzmann equation.

The aim of this paper is to describe a new method to compute the
effect of $CP$-violation coming from extensions of the SM on the
mechanism of electroweak baryogenesis and,
in these respects, it should be regarded as
alternative and complementary to the one proposed in refs.
\cite{newmethod1,newmethod2}.
The method, valid
for all wall shapes and sizes, is entirely based on
a nonequilibrium QFT diagrammatic approach and naturally
incorporates the effects of the  incoherent nature of plasma physics on
$CP$-violating observables. As such, it should be considered
as a first step towards a full
nonequilibrium quantum kinetic theory approach to electroweak baryogenesis.
For sake of clarity, the method is also illustrated through an example
in the framework of the two-Higgs doublet model.
\vspace{0.5cm}
\begin{flushleft}
{\large\bf 2. The method}
\end{flushleft}
\vspace{0.5cm}

The ordinary quantum field theory (at finite temperature), which mainly
deals with transition amplitudes in particle reactions, is not useful to
study the dynamics of classical order parameters. This is because we
need their temporal evolution with definite initial conditions and not
simply the transition amplitude of particle reactions with fixed initial
and final conditions. The most appropriate extension of the field theory
to deal with these issues it to generalize the time contour of
integration to a closed-time path (CPT). More precisely, the time integration
contour is deformed to run from $-\infty$ to $+\infty$ and back to
$-\infty$ \cite{chou}. The CPT formalism is a powerful Green's function
formulation for describing nonequilibrium phenomena in field theory. It
allows to describe phase-transition phenomena and to obtain a
self-consistent set of quantum Boltzmann equations.
The formalism yields various quantum averages of
operators evaluated in the in-state without specifying the out-state.
As with the Euclidean time formulation, scalar (fermionic) fields $\phi$ are
still periodic (anti-periodic) in time, but with
$\phi(t,\x)=\phi(t-i\beta,\x)$, $\beta=1/T$.
Temperature appears due to boundary
condition, but now time is explicitly present in the integration
contour.

As a consequence of the time contour, we must now identify field
variables with arguments on the positive or negative directional
branches of the time path. This doubling of field variables leads to
four different real-time propagators on the contour \cite{chou}
for a generic scalar field $\phi$ (and analogously for fermionic fields)
\begin{eqnarray}
G^{++}_\phi\left(x-x^\prime\right)&=&i\langle
T_{+}\phi(x)\phi(x^\prime)\rangle,\nonumber\\
G^{- -}_\phi\left(x-x^\prime\right)&=&i\langle
T_{-}\phi(x)\phi(x^\prime)\rangle,\nonumber\\
G^{+ -}_\phi\left(x-x^\prime\right)&=&i\langle
\phi(x^\prime)\phi(x)\rangle=G^{>}_\phi\left(x-x^\prime\right),\nonumber\\
G^{- +}_\phi\left(x-x^\prime\right)&=&i\langle
\phi(x)\phi(x^\prime)\rangle=G^{<}_\phi\left(x-x^\prime\right),
\end{eqnarray}
where $T_{+}$ and $T_{-}$ indicate the chronological and anti-chronological
ordering, $G^{++}_\phi$ is the usual physical (causal) propagator. The
other three propagators come as a consequence of the time contour and
are auxiliary (unphysical) propagators.

The explicit expressions for the above propagators in terms of their
momentum space Fourier transforms are given by
\begin{equation}
G_\phi\left(x-x^\prime\right)=i\:\int\:\frac{d^3 \k}{(2\pi)^3}\:{\rm e}^{
i\k\cdot\left(\x-\x^\prime\right)}\:
\left(
\begin{array}{cc}
G^{++}_\phi(\k,t-t^\prime) & G^{+ -}_\phi(\k,t-t^\prime)\\
G^{- +}_\phi(\k,t-t^\prime) & G^{- -}_\phi(\k,t-t^\prime)
\end{array}\right),
\end{equation}
where
\begin{eqnarray}
G^{++}_\phi(\k,t-t^\prime)&=&G^{>}_\phi(\k,t-t^\prime)\theta(t-t^\prime)+
G^{<}_\phi(\k,t-t^\prime)\theta(t^\prime-t),\nonumber\\
G^{- -}_\phi(\k,t-t^\prime)&=&G^{>}_\phi(\k,t-t^\prime)\theta(t^\prime-t)+
G^{<}_\phi(\k,t-t^\prime)\theta(t-t^\prime),\nonumber\\
G^{+ -}_\phi(\k,t-t^\prime)&=&G^{>}_\phi(\k,t-t^\prime),\nonumber\\
G^{- +}_\phi(\k,t-t^\prime)&=&G^{<}_\phi(\k,t-t^\prime).
\end{eqnarray}

The finite-temperature, real-time propagator
$G^{++}_\phi(\k,t-t^\prime)$ can be written in terms of the spectral
function $\rho(\k,k_0)$ \cite{ww}
\begin{equation}
G^{++}_\phi(\k,t-t^\prime)=\int_{-\infty}^{+\infty}\:
\frac{d k^0}{2\pi}\:{\rm e}^{-i k^0(t-t^\prime)}\:\rho(\k,k^0)\:\left\{
\left[1+n(k^0)\right]\theta(t-t^\prime)+n(k^0)\theta(t^\prime-t)\right\},
\label{rho}
\end{equation}
where $n(k^0)$ is the distribution function.

To account for
interactions with the surrounding particles of the thermal bath,
particles must be
substituted by quasiparticles and dressed propagators are to be adopted
(the use of
the full corrected propagators should be done with some care to avoid an
overcounting of diagrams \cite{par}).
 Self-energy
corrections at one- or two-loops to
the propagator modify the dispersion relations and
introduce nontrivial effects (damping) due to the imaginary
contributions to the self-energy: $\Sigma(k)={\rm Re}\: \Sigma(k)+i\:{\rm
Im}\:\Sigma(k)$ \cite{par}. Due to the nonvanishing ${\rm Im}\:\Sigma$,
$\rho(\k,k^0)$ acquires in the weak limit a finite width
\begin{eqnarray}
\Gamma(k)&=&-\frac{{\rm
Im}\:\Sigma(\k,\omega)}{2\:\omega(k)},\nonumber\\
\omega^2(k)&=&\k^2+m^2+{\rm Re}\:\Sigma(\k,\omega),
\end{eqnarray}
where $m$ indicates the tree-level mass, and is expressed by
\begin{equation}
\rho(\k,k^0)=i\:\left[\frac{1}{(k^0+i\:\varepsilon+ i\:\Gamma)^2-\omega^2(k)}-
\frac{1}{(k^0-i\:\varepsilon-i\:\Gamma)^2-\omega^2(k)}\right].
\end{equation}
It is easy to show that the free propagator is obtained when taking the
limit $\Gamma\rightarrow 0$.
Equation (6) has four poles in the complex plane $q^0$ plane given by $
\omega\pm i\:\Gamma$ and $-\omega\pm i\:\Gamma$.
Performing the integration over $k^0$ in the Eq. (\ref{rho}),
one gets \cite{ww}
\begin{eqnarray}
\label{a}
G_\phi^{>}(\k,t-t^\prime)&=&\frac{1}{2\:\omega}\left\{
\left[1+n(\omega-i\Gamma)\right]\:{\rm
e}^{-i(\omega-i\Gamma)(t-t^\prime)}+n(\omega+i\Gamma)\:
{\rm
e}^{-i(\omega+i\Gamma)(t-t^\prime)}\right\},\nonumber\\
G_\phi^{<}(\k,t-t^\prime)&=&G_\phi^{>}(\k,t^\prime-t).
\end{eqnarray}

In what follows we shall adopt  dressed propagators to compute the thermal
averages of composite operators, allowing us to naturally and
self-consistently
include  the effects of the  incoherent nature of plasma physics and to show
that $CP$-violating quantities vanish in the limit of fast
interactions: as already evident from expression (\ref{a}),
$\tau\simeq\Gamma^{-1}$ is a
natural temporal cut-off and any contribution at
times $t_x<t_z$ to
a $CP$-violating observable computed at the instant
$t_z$  should be expected to vanish for $(t_z-t_x)\la \tau$.

We are now in the position to compute $CP$-violating sources resulting from
particle interaction with an expanding bubble during a first-order
electroweak phase
transition.
Our starting point is the CPT
finite temperature generating functional for
the 1PI Green's functions with insertion of a composite
operator $\hat{O}(z)$ (in
the following $\hat{O}(z)$ will represent a particle current)
\cite{thicknoi}
\begin{equation}
\Gamma\left[\phi_{{\rm cl}}^\alpha,J_O\right]=
W\left[J^\alpha,J_O\right]-\sum_\beta
\:\int_c\:d^4x\: J_\beta(x)\phi_{{\rm cl}}^\beta(x),
\end{equation}
where the $\phi_{{\rm cl}}$'s are the different classical fields
of the theory with sources
$J$'s and will be  labelled with greek indices,
$J_O$ is the source for the operator $\hat{O}$ and the subscript
$c$ on the time
contour reminds that the closed-time path has been chosen \cite{chou}.
Note that the Legendre transformation
has been performed only on the fields and not on the composite
operator \cite{pok}. To avoid
any confusion, we point out that the terminology {\it classical fields}
should be understood in the QFT sense \cite{pok}, {\it i.e.} the
$\phi_{{\rm cl}}$'s are the fields which extremize the combination
\begin{equation}
\label{cc}
\Gamma\left[\phi^\alpha,J_O\right]+\sum_\beta
\:\int_c\:d^4x\: J_\beta(x)\phi^\beta(x)
\end{equation}
and from now on we will abolish the suffice ${\rm cl}$.

According to the time contour, any generic quantity $X$ is doubled to
$X_{+}$ and $X_{-}$. Although $X_{+}$ and $X_{-}$ are actually the same,
one has to regard them different from each other for technical reasons.
The association of $X_{{\rm c}}=(X_{+}+X_{-})/2$ with the physical field
imposes we take $X_\Delta=X_{+}-X_{-}=0$ at the end of the calculation
\cite{chou}.

The quantity we are interested in is the (temporal evolution of the)
expectation value $O_{{\rm
c}}(z)$
of the
operator $\hat{O}(z)$ on the background given by the fields
$\phi_{{\rm c}}^\alpha$:
\begin{eqnarray}
O_{{\rm c}}(z)&=&\frac{O_{+}(z)+O_{-}(z)}{2},\nonumber\\
O_{\pm}(z)&=&\left.\frac{1}{\pm i}
\frac{\delta\Gamma\left[\phi_{+}^\alpha,\phi_{-}^\alpha,
J_O^{\pm}\right]}{\delta J_{O}^{\pm}(z)}\right|_{J_{O}^{+}=
J_{O}^{-}=0}=
{\cal O}_{\pm}\left[\phi^\alpha_{+},\phi_{-}^\alpha\right](z).
\end{eqnarray}

The minus sign in front of the functional derivative $\delta/\delta
J_O^{-}$ takes into account that time now runs only forward.

Following a standard procedure in QFT, we can now expand
the (finite temperature) generating functional
$\Gamma\left[\phi^\alpha,J_O\right]$ in terms of the classical fields of
the theory $\phi_{\pm}^\alpha(x)$ \cite{pok}
\begin{equation}
\Gamma\left[\phi^\alpha,J_O\right]=
\sum_{n=0}^{\infty}\:
\sum_{i_1,\cdot\cdot\cdot,i_n}\frac{1}{n!}\:
\int_c\:d^4 x_1\cdot\cdot\cdot d^4 x_n\:
\Gamma^{(n)}_{i_1,\cdot\cdot\cdot,i_n}(x_1,\cdot\cdot\cdot,x_n)
\:\phi_{i_1}(x_1)\cdot\cdot\cdot\phi_{i_n}(x_n),
\end{equation}
where, with a shorthand notation, we have
indicated through the indices $i_1,\cdot\cdot\cdot i_n$ all the
possible combinations of the scalar fields of the theory
$\phi^\alpha_{\pm}$, and the coefficients of the expansion
\begin{equation}
\Gamma^{(n)}_{i_1,\cdot\cdot\cdot,i_n}(x_1,\cdot\cdot\cdot,x_n)=
\left.\frac{\delta^{(n)}\Gamma\left[\phi_{+}^\alpha,\phi_{-}^\alpha,
J_O^{\pm}\right]}{\delta\phi_{i_1}(x_1)\cdot\cdot\cdot
\delta \phi_{i_n}(x_n)}\right|_{\phi^{\alpha}_{\pm}=0}
\end{equation}
are the $n$-point 1PI Green's functions computed for vanishing
$\phi^\alpha_{\pm}(x)$.

The functional ${\cal
O}_{\pm}\left[\phi^\alpha_{+},\phi_{-}^\alpha\right](z)$ can then
be written
in a power series of
$\phi^\alpha_{\pm}$
\begin{equation}
\label{exp}
{\cal O}_{\pm}\left[\phi^\alpha_{+},\phi^\alpha_{-}
\right](z)=\sum_{n=0}^{\infty}\:
\sum_{i_1,\cdot\cdot\cdot,i_n}\frac{1}{n!}\:
\int_c\:d^4 x_1\cdot\cdot\cdot d^4 x_n\:
{\cal O}^{(n)}_{i_1,\cdot\cdot\cdot,i_n}(x_1,\cdot\cdot\cdot,x_n;z)
\:\phi_{i_1}(x_1)\cdot\cdot\cdot\phi_{i_n}(x_n).
\end{equation}
The coefficients of the expansion (\ref{exp}) are the $n$-point
1PI Green's functions with one insertion of the operator $\hat{O}_{\pm}$
computed for vanishing
$\phi^\alpha_{\pm}(x)$
\begin{equation}
{\cal O}^{(n)}_{i_1,\cdot\cdot\cdot,i_n}(x_1,\cdot\cdot\cdot,x_n;z)=
\left.\frac{1}{\pm
i}\frac{\delta^{(n+1)}\Gamma\left[\phi_{+}^\alpha,\phi_{-}^\alpha,
J_O^{\pm}\right]}{\delta\phi_{i_1}(x_1)\cdot\cdot\cdot
\delta \phi_{i_n}(x_n)\delta J_O^{\pm}(z)}\right|_{\phi^{\alpha}_{\pm}=
J_{O}^{+}=
J_{O}^{-}=0}
\end{equation}
and can be used to compute through a  diagrammatic approach
the temporal evolution of the vacuum expectation value $O_{{\rm c}}(z)$
of the operator $\hat{O}(z)$ in a generic space-time dependent
background described by the classical fields $\phi_{\pm}^\alpha(x)$. For our
practical purposes, such a background will be the bubble wall configuration
expanding in the thermal bath with velocity $v_{{\rm w}}$, which
extremizes the combination (\ref{cc}) with
vanishing sources.

Thus, given a certain theory with the necessary amount of
$CP$-violation and predicting
nonequilibrium conditions at a certain space-time
point through the passage of the bubble wall, the
general formulae (10-14) can be
applied to compute in the rest
frame of the thermal bath and at a certain point $\z$ the temporal evolution
of the nonequilibrium expectation
values $J_{{\rm c}}^\mu(t_z,\z)$
of the  current operators $J^\mu(t_z,\z)$ for the different particles
of the theory.

Even if the general formalism described so far might
seem rather cumbersome, we shall see through a
practical example that it leads to the computation of
physical quantities whose meaning  is clear and well-established,
making the
physical interpretation of the whole picture rather intuitive. Moreover,
the expansion (13) is adequate to demonstrate the physics
of quantum interference required to generate $CP$-violating observables.
It originates a series of powers in the classical Higgs fields whose
imaginary parts leads to nonvanishing $CP$-violating observables.
$CP$-violating
sources $\gamma_Q(z)$
(per unit volume and unit time)
of a generic charge
$Q$ associated to the current $J^\mu(z)$
and accumulated by the moving wall at a point
$\z$ of the plasma can then be constructed from
$J_{{\rm c}}^\mu(z)$ \cite{newmethod1,newmethod2}
\begin{equation}
\gamma_Q(z)=\partial_\mu J_{{\rm c}}^\mu(z).
\end{equation}
We point out that this definition of the $CP$-violating source
$\gamma_Q(z)$, though adequate to describe the damping effects
originated by the
plasma interactions, does {\it not} involve any relaxation time scale
arising when diffusion and particle changing interactions are included.
In this paper we will focus on the computation of $CP$-violating
observables through a nonequilibrium QFT approach
and take the same point of view
as in  refs. \cite{newmethod1,newmethod2}: one can
leave aside diffusion and particle changing interactions
 and account for them
independently in the rate equations. It is clear, however,  that a full
nonequilbrium QFT approach to electroweak baryogenesis, based on a
complete set
of quantum Boltzmann equations, should describe consistently all the
effects. Indeed, the CPT formalism, in combination with the so-called
non-local source theory and loop expansion techniques developed by
Cornwall, Jackiw and Tamboulis \cite{cjt}, could provide the
Ginzburg-Landau equation \cite{gin} for the order parameter (in our case the
Higgs background field) and the
generalized Dyson-Schwinger equations which incorporate the initial
state correlations and provide a systematic treatment of the quantum
correlations to any order of perturbation theory. This procedure would
yield two distinct equations for each of the Wigner functions: the
renormalization equation and the transport equation. The latter
should account in a
self-consistent way
for all the effects (diffusion, $CP$-violation, damping, particle changing
interactions, {\it etc.}) giving rise to the final baryon asymmetry.
This analysis is, however, beyond the scope of the present work.

Due to their generality, expressions (10-14) are valid for a generic wall
shape and size
$L_{{\rm w}}$ and are not based on any assumptions on the relative
magnitude of the mean free paths and thickness of the bubble wall. In
this respect, the method described in this paper should be regarded as
an extension of the linear response method \cite{pit}
(as accurate as the thick wall approximation is made)
to compute
the charge current densities produced from an initial $CP$-symmetric
thermal particle distribution when space-time dependent
$CP$-violating terms are turned on in the Hamiltonian for a time equal
to the thermalization time.
\vspace{0.5cm}
\begin{flushleft}
{\large\bf 3. An example of construction of a $CP$-violating source}
\end{flushleft}
\vspace{0.5cm}

Let us now illustrate the method by means of a specific situation:
the two-Higgs doublet model;
more in particular, we will work out the $CP$-violating
source constructed from the Higgs current operator $J^\mu_{1}(z)$ associated
to the neutral Higgs field $H_1^0$
\begin{equation}
J_1^\mu(z)= i\left(H_1^\dagger\partial^\mu H_1-\partial^\mu
H_1^\dagger H_1\right).
\end{equation}
The reader should keep in mind that the method applies to
other theories as well.

The most general tree level scalar potential for the two-Higgs doublet
model is given by
\begin{eqnarray}
\label{pot}
V&=& m_1^2\left|H_1\right|^2+m_2^2\left|H_2\right|^2-\left(m_3^2 H_1
H_2+{\rm h.c.}\right)+\lambda_1 \left|H_1\right|^4
+\lambda_2 \left|H_2\right|^4\nonumber\\
&+&\lambda_3 \left|H_1\right|^2\left|H_2\right|^2
+\lambda_4\left|H_1 H_2\right|^2+\left[\lambda_5\left(H_1 H_2\right)^2+
\lambda_6\left|H_1\right|^2H_1 H_2+\lambda_7\left|H_2\right|^2H_1 H_2
+{\rm h.c.}\right].
\end{eqnarray}
Of the two phases of the classical vacuum
expectation values of the Higgs fields $\delta_1$ and $\delta_2$, defined by
$H_{{\rm c},i}(x)=v_i(x)\:{\rm exp}\left[i\delta_i(x)\right]$
$(i=1,2)$,
only the gauge invariant combination $\theta=\delta_1+\delta_2$ appears
in the scalar potential, whereas the orthogonal combination represents
the gauge phase. We assume that the parameters of the Lagrangian are
such that, when loop corrections are considered, the potential assumes a
double-well shape and expanding bubble wall solution exists described by
the configuration $H_{{\rm c},i}(x)$.

The interesting dynamics for baryogenesis takes place in a region
close to or inside the bubble wall and we approximate it with an infinite plane
traveling at a constant speed $v_{{\rm w}}$ along the $z$-axis.

Since the coefficients of the expansion (13) must be computed
for vanishing $\phi^\alpha_{\pm}(x)$,
we must deal with resummation of the propagators of the
Higgs fields in order to deal with infrared divergencies \cite{arnold}.
In the unbroken phase, the Higgs spectrum contains two complex
electrically neutral fields and two charged ones. At the tree-level, the
squared masses of one of the neutral states and one of the charged ones
are negative, since the origin of the field space becomes a minimum of
the effective potential only after inclusion of the finite temperature
corrections. The resummation can be achieved by considering the
propagators for the eigenstates of the thermal mass matrix, which has
positive eigenvalues given by
\begin{equation}
M_{h,H}^2(T)=\frac{m_1^2(T)+m_2^2(T)\mp\sqrt{\left(
m_1^2(T)-m_2^2(T)\right)^2+4\:m_3^4(T)}}{2},
\end{equation}
where the $m_i^2(T)$ are the thermal corrected mass parameters of the
potential (\ref{pot}), $m_1^2(T)\simeq m_1^2+(3/16)g^2 T^2$,
$m_2^2(T)\simeq m_2^2+(1/4)h_t^2 T^2$, while $m_3^2(T)$ receives only
logarithmic corrections in $T$ \cite{noi}. Here $g$ and $h_t$ are the $SU(2)_L$
gauge coupling and the top Yukawa coupling, respectively.
Correspondingly, the neutral complex eigenstates are given by
\begin{equation}
\left\{
\begin{array}{ccc}
h&=& \cos\beta_{hH}\:H_1^0 + \sin\beta_{hH}\:H_2^{0 *},\\
H&=& -\sin\beta_{hH}\:H_1^0 + \cos\beta_{hH}\:H_2^{0 *},
\end{array}
\right.
\end{equation}
where
\begin{equation}
\sin 2\beta_{hH}=\frac{2\:m_3^2(T)}{M_h^2(T)-M_H^2(T)}.
\end{equation}
Completely analogous formulae hold for the charged eigenstates.

Given the classical action (remind that time runs only forward)
\begin{equation}
S\left[H_1,H_2\right]=\int\:d^4x\left\{{\cal
L}\left[H_{1,+},H_{2,+}\right]-{\cal
L}\left[H_{1,-},H_{2,-}\right]\right\},
\end{equation}
it is not difficult to show that the the vacuum
expectation value $J_{1,{\rm c}}^\mu(z)$ gets (beyond the tree level ones)
contributions
from four different one-loop Feynman diagrams, symbolically
depicted in Fig. 1,
which are obtained by assigning
in all the possible ways the  space-time points $x$ and $z$
 on the positive
or negative time branches. One gets
\begin{equation}
J_{1,{\rm c}}^\mu(z)=\frac{J_{1,+}^\mu(z)+J_{1,-}^\mu(z)}{2}=
-4\:\lambda_5\:\sin 2\beta_{hH}\:\int\:d^4 x\:H(x)\:\Sigma^\mu(x,z)+
{\cal O}\left[(H_{{\rm c}}/T)^4\right],
\end{equation}
where we have made an expansion in the parameter $(H_{{\rm c}}/T)$. The
function
\begin{equation}
\Sigma^\mu(x,z)= \theta(t_z-t_x)\:\lim_{y\rightarrow z}\:\left\{
\partial^\mu_y\:
{\rm Im}\left[G^{++}_{H}(y-x)\:
G^{++}_{h}(x-z)\right]\right\}-(h\leftrightarrow H)
\end{equation}
is expressed in terms of the {\it retarded} self-energy Green's function
and we have defined
\begin{equation}
H(x)={\rm Im}\left(
H_{1,{\rm c}}^0\:H_{2,{\rm
c}}^0\right)(x).
\end{equation}
Notice that, being  dependent on $H(x)$,
$J_{1,{\rm c}}^\mu$
is vanishing if no $CP$-violation is
present in the scalar sector.

Using the spectral representation (6), we can cast
the function $\Sigma^\mu(x,z)$ into the form
\begin{eqnarray}
\Sigma^\mu(x,z)&=&-i\:\theta(t_z-t_x)\:\int\:\frac{d^4 q_1}{(2\pi)^4}\:
\int\:\frac{d^4 q_2}{(2\pi)^4}\:q_1^\mu\:\rho_H\left(q_1\right)
\:\rho_h\left(q_2\right)\nonumber\\
&\times &\left[n_h(q^0_2)-n_H(q^0_1)\right]\:
{\rm e}^{-i(z-x)\cdot(q_1-q_2)}-(h\leftrightarrow H).
\end{eqnarray}
Performing the $q_1^0$ and $q_2^0$ integrations, one can easily show that
$J_{1,{\rm c}}^\mu(z)$ is vanishing in the limit of {\it constant}
space-time $CP$-violating background $H(x)$: this is the very
well-known result that $CP$-violating sources arise only in a
nontrivially space-time dependent $CP$-violating Higgs background.

The presence of the retarded Green's function carrying the
$\theta(t_z-t_x)$ makes the physical interpretation of expression (22)
rather intuitive: let us imagine the bubble wall configuration moving
along the $z$-axis and
divided into many strips along the same axis; when each of these strips
passes from
$t_x$ to $t_x+dt_x$
through a small volume element centered at the
point $\z$ in the thermal bath, it deposits there
a certain amount of charge and the vacuum expectation value of
$J_{1}^\mu$ at the time $t_z$ turns out to be the sum of all these
contributions received for times $t_x<t_z$. Moreover, because of the
presence of a finite damping rate $\tau^{-1}$,
the strips' contributions  are not all equivalent, but they will be
weighted by the exponential factors present in the Green's functions,
thus
making the strips'contributions negligible
for times $(t_z-t_x)\ga \Gamma^{-1}$.

The advantage of the method illustrated in this paper is that very
general formulae, like Eqs. (22-23), can be obtained for the temporal
evolution of $CP$-violating observables without any
particular assumption on the relative magnitude of the mean free paths
and the thickness of the wall, and on the particle distribution functions
(eventually determined by the Boltzmann equations of motion).
The standard thin wall and thick wall
limits are recovered in the limit $\Gamma L_{{\rm w}}\rightarrow 0$
and $\infty$, respectively, once the Green's functions (7) are inserted
in the expression (23) for $\Sigma^\mu(x,z)$.

In order to deal with analytic expressions, we can work out
the thick wall limit and simplify expressions
(22-24) by performing a derivative expansion
\begin{equation}
H(x)\simeq H(z)+\partial_x^\nu \left. H\right|_{x=z}
\left(x_\nu-z_\nu\right)+{\cal O}
\left[\left(\Gamma_{h,H}L_{{\rm w}}\right)^2\right].
\end{equation}
This expansion is valid only when the mean free paths $\tau\simeq
\Gamma^{-1}$ are smaller than the scale of variation of the Higgs
background determined by the wall thickness $L_{{\rm w}}$. An estimate
of the Higgs damping rate in the Standard Model was obtained in ref.
\cite{damping} in the low momentum limit and can be used here only to
give
very crude estimate of the Higgs coherent times, $\tau^{-1}\sim
(10^{-2}-10^{-1})\:T$ (with such values, our derivative expansion is not
perfectly justified since the wall thickness can span the range
$(10-100)/T$. Its benefit relies on the possibility to obtain
analytical expressions in $\Gamma$ and
we adopt it for pedagogical purposes).

To work out exactly the vacuum expectation value
of the Higgs current $J_{1,{\rm
c}}^\mu(z)$ one should know the exact form of the distribution
functions
which, in ultimate analysis, is provided by solving the Boltzmann equations.
However, any departure from thermal equilibrium distribution functions
is caused at a given point by the passage of the wall and, therefore,
is  ${\cal O}(v_{{\rm
w}})$.
Working with thermal equilibrium distribution
functions in Eq. (7) amounts to ignoring terms of order $v_{{\rm
w}}^2\la 1$, which is as accurate as the bubble wall is moving slowly in
the plasma and we shall adopt this approximation from now on.
At this order, the $G_\phi^{++}(\k,t-t^\prime)$
Green's function
becomes
\begin{eqnarray}
G_\phi^{++}(\k,t-t^\prime)&=&\frac{{\rm e}^{-\Gamma\left|t-t^\prime\right|}}{
2\omega\left[{\rm cosh}(\beta\omega)-{\rm
cos}(\beta\Gamma)\right]}\left\{{\rm sinh}(\beta\omega){\rm cos}(\omega
\left|t-t^\prime\right|)\right.\nonumber\\
&+&\left.{\rm sin}(\beta\Gamma){\rm sin}(\omega
\left|t-t^\prime\right|)+i\left[{\rm cos}(\beta\Gamma)-
{\rm cosh}(\beta\Gamma)\right]{\rm sin}(\omega
\left|t-t^\prime\right|)\right\}.
\end{eqnarray}
To demonstrate the conflict between the incoherent nature of
fast plasma interactions and the
coherent phenomenon of $CP$-violation, we can take the ideal limit
$\Gamma_{h}\simeq\Gamma_{H}\gg T$. Even if the physical lower limit
in the temporal integration is
about $t_z-L_{{\rm w}}$, we can safely take it $-\infty$ in the thick
wall approximation.
Using expressions (22-27) one can show
that
\begin{eqnarray}
\label{clear}
\gamma_1(z)&=&\partial_\mu
J_{1,{\rm c}}^\mu(z)\simeq
-2 \lambda_5 m_3^2\:\partial_{t_z}H(z)\:\int_{-\infty}^{t_z}\:dt_x\:
\left(t_z-t_x\right)\:\Gamma_h^2\:{\rm e}^{-2\Gamma_h(t_z-t_x)}
\:
\int\:\frac{d^3\k}{(2\pi)^3}\:\frac{\sin(\beta\Gamma_h)}
{\omega_h^4}\nonumber\\
&\times&\left[\cos(\omega_h+\omega_H)(t_z-t_x)
-\cos(\omega_h-\omega_H)(t_z-t_x)\right]
+
{\cal O}\left[v_{{\rm w}}^2,\left(\Gamma_h L_{{\rm w}}\right)^2,
\left(M_{h,H}/T\right)^4\right].
\end{eqnarray}
Performing the integration in $t_x$ one gets
\begin{equation}
\gamma_1(z)\simeq
 \lambda_5\:
\frac{\sin(\beta\Gamma_h)}{2\:\pi^2}\:\left(\frac{
T\:m_3^2}{\Gamma_h^2}\right)\:\partial_{t_z}
H(z)\:{\cal I}_h\left(T,\Gamma_h\right)+
{\cal O}\left[v_{{\rm w}}^2,\left(\Gamma_h L_{{\rm w}}\right)^2,
\left(M_{h,H}/T\right)^4\right],
\end{equation}
where
\begin{eqnarray}
{\cal I}_h\left(T,\Gamma_h\right)&=&
\int_0^{\infty}dx\:\frac{x^2}{(x^2+y^2)}\:\frac{1}{
{\rm cosh}(\sqrt{x^2+y^2})-{\rm cos}(\beta\Gamma_h)}\nonumber\\
&\simeq&
\frac{2}{\sin(\beta\Gamma_h)}\:{\rm arctg}\left(\frac{
\sin\beta\Gamma_h}{1-\cos\beta\Gamma_h}\right)+{\cal O}(y^2),\nonumber\\
y&=&M_h(T)/T.
\end{eqnarray}
The steep dependence of $\gamma_1(z)$ on $\Gamma_h$ reflects the high
suppression coming from decoherence due to incoherent scatterings in
the plasma whose rate increases with $\Gamma_h$. The origin of this
strong suppression is similar to the one forbidding
electroweak baryogenesis in SM \cite{huet} and is
evident from Eq. (\ref{clear}):
due to the
exponential cut-off ${\rm exp}\left[-\Gamma_h t\right]$,
as the wall crosses a small volume
element of the plasma the current density deposited
at the instant $t_z$ into it  may receive relevant contributions only
from those sheets of the bubble wall which pass through the volume
element
at times
in the range $\left[t_z-\tau_h,t_z\right]$, which vanishes for
very small interaction times $\tau_h$.

In the opposite limit of very weak interactions $\Gamma_h\la T$, a
straightforward calculation leads to
\begin{equation}
\gamma_1(z)\simeq
\frac{\lambda_5}{32\:\pi}\Gamma_h\:T\:\frac{m_3^2}{M^3_h(T)}
\partial_{t_z}H(z)+
{\cal O}\left[v_{{\rm w}}^2,\left(\Gamma_h L_{{\rm w}}\right)^2,
\left(M_{h,H}/T\right)^4\right].
\end{equation}
Eq. (31) warrants some comments. First, we notice that the
momentum integration giving rise to the $1/M_h^3$ dependence in the
above expression
is
infrared dominated: quasiparticles with long wavelengths and
incident perpendicularly to the wall give a large contribution to
$\gamma_1(z)$ and a classical approximation is not adequate to
describe the quantum interference nature of $CP$-violation.
Secondly, the vanishing  of $\gamma_1(z)$ for very
small interaction rates
$\Gamma_h\ll T$ is in agreement with what found in ref.
\cite{newmethod1} and should be expected since the propagation is
semi-classical in that limit (even if one should not
expect to recover our result by a complete classical treatment, because
the momentum
integral is dominated by particles with long wavelengths for which a
full classical approach  is inappropriate).
Working in the thick wall
limit and with very small damping rates makes (contrary
to what happens for strong damping rates) all the different
sheets of the bubble wall equivalent. In these limits, there are no
time-scales involved and the amount of charge
deposited in a small volume element turns out to be approximately
independent of time, resulting in a vanishing $\gamma_1(z)$.
As a matter of fact, such a behaviour for
very small damping rates
should not be  completely trusted since our derivative expansion
may be applied only for  $\Gamma_h\ga L_{{\rm w}}^{-1}$.
Also, as we mentioned above,
our computation tacitly assumes   large relaxation times
provided by the diffusion properties of the particles in the plasma.
When considering the effects of finite relaxation times,
the behaviour of $\gamma_1(z)$ for $\Gamma_h \ll T$ should
differ from the one obtained here
since particles are allowed to diffuse away from the
region centered at the point $\z$.
\vspace{0.5cm}
\begin{flushleft}
{\large\bf 4. Conclusions}
\end{flushleft}
\vspace{0.5cm}

In this paper we have proposed a general method to compute
$CP$-violating observables in the context of electroweak baryogenesis.
It is valid for all shapes of the expanding bubble wall and relies on
a nonequilibrium quantum field theory diagrammatic approach. Dressed
propagators have been adopted to account for the interactions with the
surrounding particles, which modify the dispersion relation and introduce
damping effects. The suppression of $CP$-violating sources resulting
from fast incoherent thermal processes has been recovered in a direct
way. The method is complementary to the one introduced in refs.
\cite{newmethod1,newmethod2} and reproduce qualitatively their results.
Our computation based on nonequilibrium QFT
 should be regarded as a first step towards a complete understanding of the
dynamics of the electroweak
baryogenesis by means of a complete
nonequilibrium quantum kinetic theory approach. Indeed, only the latter
should be  able to properly describe the kinematics of
quasiparticle excitations in the infrared energy
region, {\it i.e.} excitations
whose Compton wavelength is larger both than  the typical interaction time and
the bubble wall width. For such quasipartilces,
the classical approximation to the Boltzmann equation is  not adequate.
We hope to address this problem in the next future.
\acknowledgments

It is a pleasure to thank A.E. Nelson for reading
the
manuscript and very useful discussions. M. Pietroni is also acknowledged
for criticisms.

\newpage
{\large\bf Figure Captions}
\vspace{1 cm}

Fig. 1: The 1-loop contribution to the vacuum expectation value of
the Higgs current $J_1^\mu(z)$.

\end{document}